\documentclass[journal]{IEEEtran}
\IEEEoverridecommandlockouts
\usepackage{cite}
\usepackage{amsmath,amsthm,amssymb,amsfonts}

\usepackage{graphicx}
\usepackage{textcomp}
\usepackage{xcolor}
\def\BibTeX{{\rm B\kern-.05em{\sc i\kern-.025em b}\kern-.08em
    T\kern-.1667em\lower.7ex\hbox{E}\kern-.125emX}}
\usepackage{multirow}
\usepackage{tabularx}
\usepackage{longtable}
\usepackage{siunitx}
\usepackage{dblfloatfix}
\usepackage{float}
\usepackage{xcolor}
\usepackage[font=footnotesize]{caption}

\DeclareCaptionLabelSeparator{periodspace}{.\quad}
\usepackage{cite}
\usepackage{multirow}
\usepackage{graphicx}
\usepackage{amsmath}
\usepackage{multicol}
\usepackage{mathtools}
\usepackage{tabularx}
\usepackage{longtable}
\usepackage{mdwlist}
\usepackage{graphicx}
\usepackage{algpseudocode}
\usepackage{balance}
\usepackage[square,numbers]{natbib}
\usepackage{booktabs}
\usepackage{url}
\usepackage{gensymb}
\usepackage{inputenc}
\usepackage{caption}
\usepackage{subcaption}
\usepackage[toc, acronym, nopostdot, nonumberlist]{glossaries}
    \makeglossaries

\newacronym{ev}{EV}{Electric Vehicle}
\newacronym{pv}{PV}{Photovoltaic}
\newacronym{thd}{THD}{Total Harmonic Distortion}
\newacronym{ups}{UPS}{Uninterrupted Power Supply}
\newacronym{vfd}{VFD}{Variable Frequency Drive}
\newacronym{pq}{PQ}{Power Quality}
\newacronym{epri}{EPRI}{Electric Power Research Institute}
\newacronym{pcc}{PCC}{Point of Common Coupling}
\newacronym{redd}{REDD}{Reference Energy Disaggregation Data}
\newacronym{hhg}{HHG}{Higher-Order Harmonics}

\begin{document}
\captionsetup[figure]{labelsep=colon,name={Fig.}}
    \title{On the Impact of High-Order Harmonic Generation in Electrical Distribution Systems \\
\thanks{This work was supported by the Sensors and Data Analytics Program of the U.S. Department of Energy Office of Electricity, under Contract No. DE-AC05-76RL01830.}
}
\author{\IEEEauthorblockN{Aaqib Peerzada, Bhaskar Mitra, Soumya Kundu and James P. Ogle}\\
\IEEEauthorblockA{\textit{Electricity Infrastructure and Buildings Division} \\
Pacific Northwest National Laboratory\\
Richland, WA 99354, USA \\
Email: \{aaqib.peerzada, bhaskar.mitra, soumya.kundu, james.ogle\}@pnnl.gov}}
\maketitle
\begin{abstract}
The modern power grid has seen a rise in the integration of non-linear loads, presenting a significant concern for operators. These loads introduce unwanted harmonics, leading to potential issues such as overheating and improper functioning of circuit breakers. In pursuing a more sustainable grid, the adoption of electric vehicles (EVs) and photovoltaic (PV) systems in residential networks has increased. Understanding and examining the effects of high-order harmonic frequencies beyond $1.5$ kHz is crucial to understanding their impact on the operation and planning of electrical distribution systems under varying nonlinear loading conditions. This study investigates a diverse set of critical power electronic loads within a household modeled using PSCAD/EMTdc, analyzing their unique harmonic spectra. This information is utilized to run the time-series harmonic analysis program in OpenDSS on a modified IEEE 34 bus test system model. The impact of high-order harmonics is quantified using metrics that evaluate total harmonic distortion (THD), transformer harmonic-driven eddy current loss component, and propagation of harmonics from the source to the substation transformer.  

% The various scenarios have been studied with increasing EV and PV penetrations. The load mix is chosen such that the effect of the harmonics are discussed along with the effective transformer loading. Our findings reveal that since PV and EV inverters are well regulated their individual  to a power system network are low however their interaction could have a substractive effect on the performance of residential transformers. 
\end{abstract}


% \printglossaries

% \printnomenclature
    \section{Introduction}

In 1992, the nonlinear load accounted for about $15$-$20\%$ of the total load served by the electric utilities in the United States. The use of nonlinear loads has steadily increased and even accelerated over the past decade, and based on a recent projection by \gls{epri}, the nonlinear load will account for roughly $50$-$70\%$ of the total utility load in the year 2024. Such rapid adoption of power electronic loads including uninterrupted power supply devices, personal computers, laptops, \gls{ev} chargers, \gls{pv} inverters, \gls{vfd}  present serious power quality concerns. The non-sinusoidal currents drawn by the nonlinear loads interact with the network impedance resulting in non-sinusoidal voltage drop \cite{grady2012understanding}. The non-sinusoidal voltage and current components are integer multiples of the fundamental component called \textit{“harmonics”}. The deterioration of the supply voltage creates stress on the electrical equipment and can potentially damage it, resulting in increased operating costs and downtime \cite{mitra2023}. Furthermore, the distorted voltage and current waveforms affect the power factor and increase the power loss in the network. Previous studies have reported many challenges caused by the harmonics such as overheating the cables and transformers and reduced equipment efficiency \cite{McLorn2017}.

In electrical distribution systems, the problem of harmonic pollution is exacerbated by the use of power factor correction capacitor banks which make phenomena like harmonic resonance much more likely. The resonance curve of a typical distribution feeder is often broad-shaped which renders the feeder sensitive to a range of harmonic orders \cite{grady2012understanding}. Moreover, it is common to have multiple capacitor banks installed over the length of the distribution feeder for power factor and voltage control applications. Hence, computer simulations are required to accurately assess the power system behavior at frequencies at or near the resonant frequency. The interaction of the harmonics injected by the nonlinear loads with the system resonant frequency can amplify the voltage and current distortion. Owing to the rapid increase in the use of nonlinear loads, analyzing the power system response to harmonic injections occurring at and beyond $1.5$ kHz is imperative. \par 
According to IEEE-519 Standard, there are strict regulations of the amount of \gls{thd} that \gls{vfd}s, \gls{pv}s and \gls{ev}s can inject at a bus \cite{ieee-519}. However, due to the difference in nature of \gls{vfd}s, \gls{ev}, and \gls{pv} operations, their interactions can create predominant high-order harmonics that are usually ignored while analyzing the system response. For allowable harmonic injections, several standards have been established \cite{en50160}, however, most discuss regulations up to the $25^{th}$ harmonic.
Current models lack the fidelity needed to capture the nonlinear behavior of loads. Furthermore, \textit{'ZIP'} based harmonic load models do not faithfully reproduce the observed harmonic spectrum of various nonlinear devices \cite{collin_component-based_2010}. Thus it becomes essential to develop models that can create harmonic-rich current/voltage datasets using detailed models that can be utilized to understand the operation of various distributed energy resources (DERs). To our best knowledge, only one such real-world dataset, referred to as Reference Energy Disaggregation Data (REDD) exists \gls{redd} \cite{redd}. However, the REDD data does not explicitly define the load clusters and it is challenging to understand the harmonic effect created by different load compositions \cite{redd}.

        In this paper, we design a framework to quantify the interaction between system resonance and high order ($\geq$ $25^{th}$) harmonics using diverse power electronic loads. Specifically, we (I) study the voltage distortion at the point of common coupling using quasi-static time-series \gls{thd} computations, (II)  evaluate the harmonic-driven eddy current loss component in a split-phase center-tapped transformer used for connecting the nonlinear loads with the rest of the distribution system and (III) demonstrate the propagation of harmonics using multiple dispersed sources of harmonic injection at locations on the distribution system with resonant frequencies equal or close to the predominant high order ($\geq 25^{th}$) harmonic frequency in the harmonic spectrum of the nonlinear load.

The rest of the paper has been organized as follows. Section \ref{modeling} discusses the PSCAD/EMTdc modeling approach to generate the high-order harmonic data for different load compositions. Section \ref{gridimpacts} discusses the phenomenon of harmonic resonance, harmonic power flow technique, and the DER models used for \gls{thd} and eddy current loss evaluations. Section \ref{Results} discusses the simulation results, and Section \ref{conc} concludes the paper with major findings and proposed future research directions.

\section{Harmonic Data Generation} \label{modeling}
\subsection{System Description}
\begin{figure}
    \centering
    \includegraphics[width=1\columnwidth]{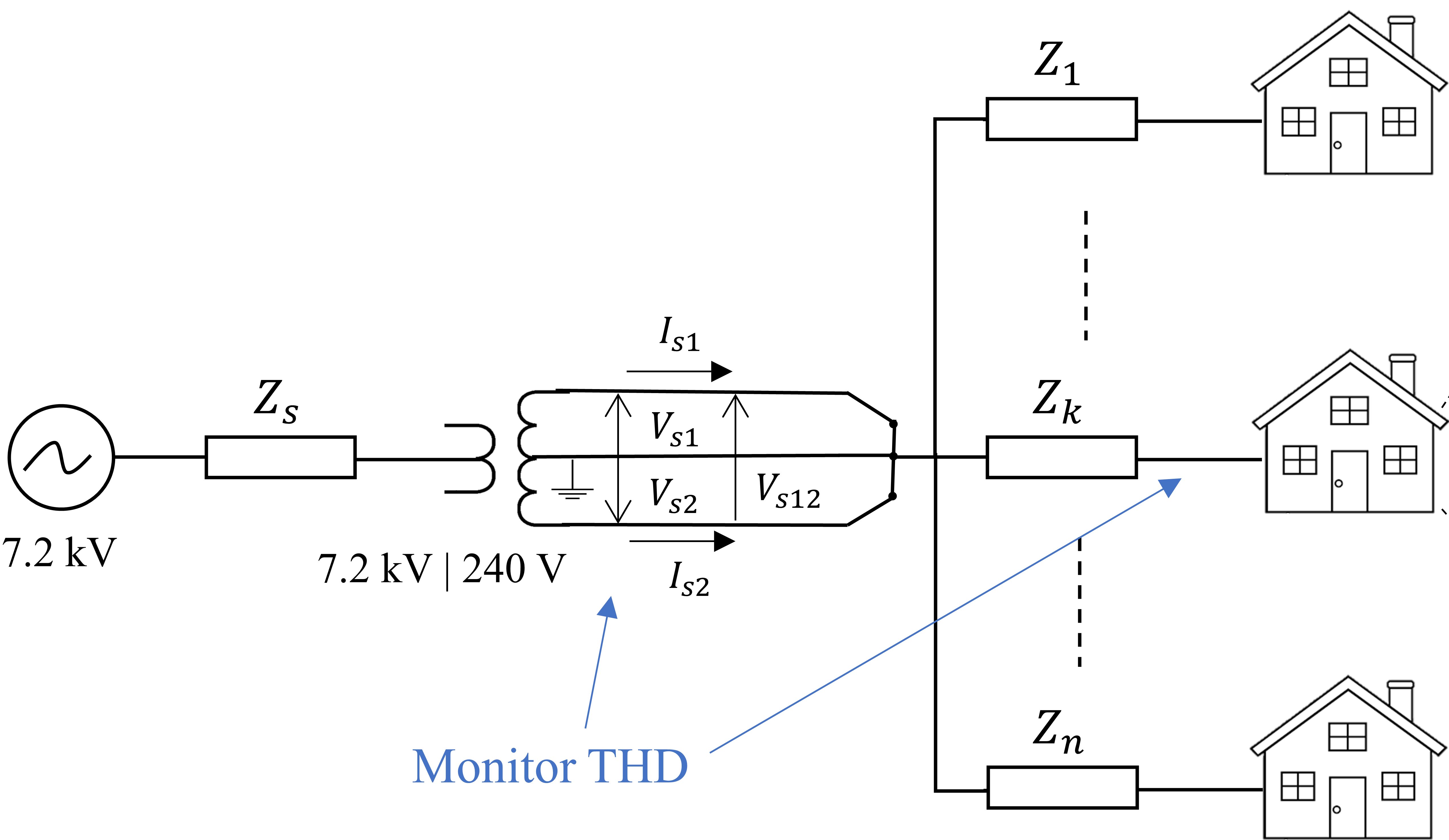}
    \caption{Simulation setup as modeled in PSCAD.}
    \label{fig:house_combo}
\end{figure}

Usually, the residential customers are supplied through a single split-phase connection in the USA. In this work, we are modeling 5 houses that have been connected to a 7.2kV/240V distribution transformer, as shown in Fig. \ref{fig:house_combo}.
Each home is comprised of four power electronic load combinations shown in Table \ref{tab:load_models} and discussed in detail in \cite{ankit2022}. 

\begin{table}[H]
\renewcommand{\arraystretch}{1.2}
\caption{Power electronics-based load models representing house appliances }
\label{tab:load_models}
\centering
\begin{tabular}{l l}
\hline
 Load model & House appliances \\
\hline
Rectifier + Buck dc-dc converter & Desktop, home entertainment\\
Rectifier + Flyback dc--dc converter & Laptop charger \\
\gls{vfd} + Induction motor & HVAC, washer, dryer \\
Boost converter + inverter & \gls{pv} system, \gls{ev} charger\\
\hline
\end{tabular}
\vspace{-3mm}
\end{table}

\begin{table}[]
\begin{tabular}{ll}
\hline
Load model                                   & Example of house appliances                                         \\ \hline
\gls{vfd} + Induction motor & HVAC, washer, dryer                                                \\
Boost converter + inverter                   & \gls{pv} system, \gls{ev} charger \\ \hline
\end{tabular}
\end{table}
\subsection{Data Generation}
\begin{figure}
    \centering
    \includegraphics[width=1\columnwidth]{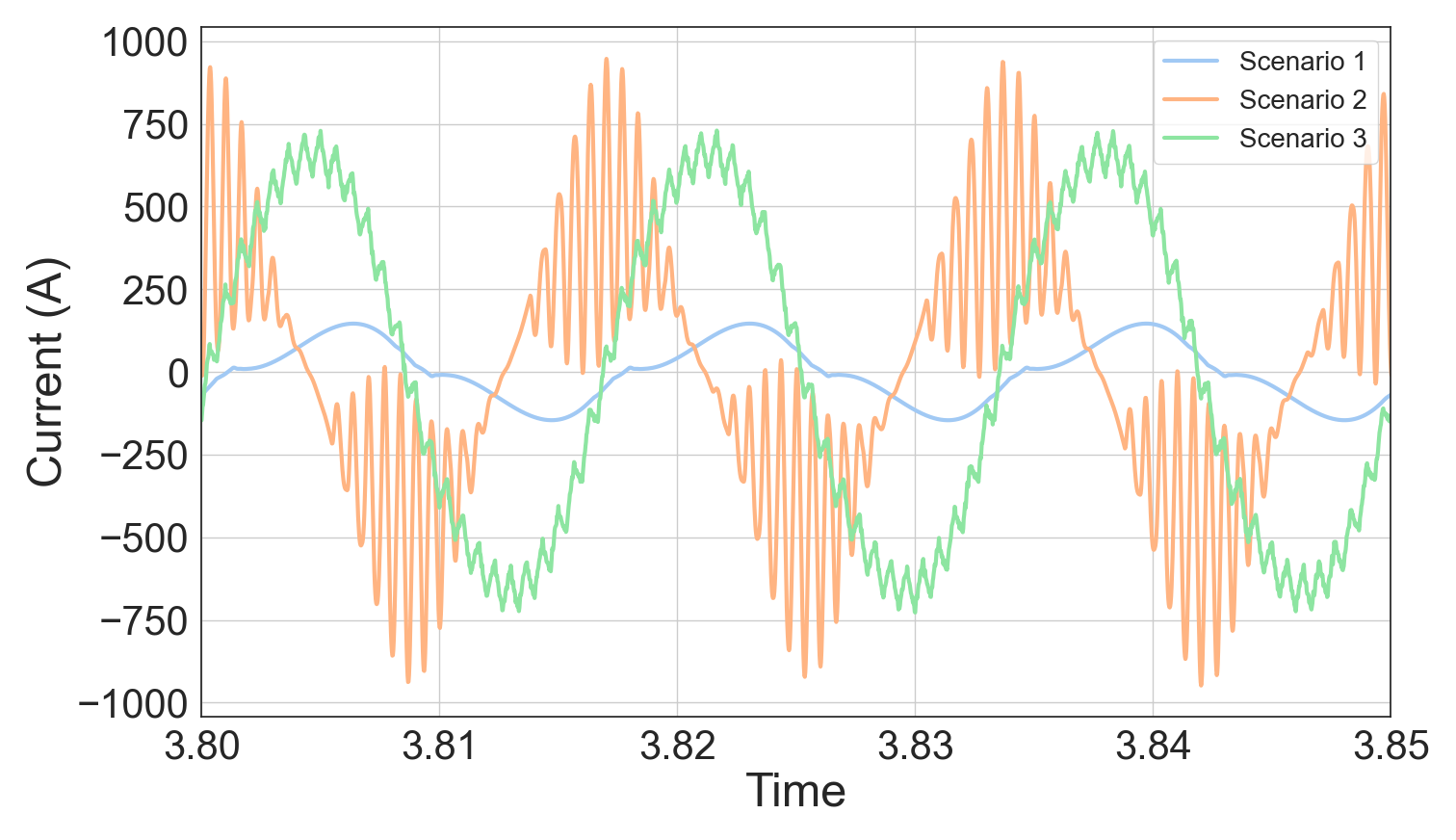}
    \caption{Current data recorded at the transformer secondary for the three scenarios}
    \label{fig:trans}
\end{figure}

The power electronic load combinations were modeled using PSCAD/EMTdc. The steady-state values of current were recorded at the secondary of the distribution transformer as shown in Fig. \ref{fig:house_combo}. A similar process was repeated for all load combinations, and three scenarios were selected where we highlight the effects of \gls{pv} and \gls{ev} penetration in residential feeders. Current and voltage data were collected from the transformer secondary at $20$ kHz to be utilized for a daily simulation of 24 hours in a modified IEEE 34 bus feeder in OpenDSS as shown in Fig. \ref{fig:trans}.
\section{Grid Impact of High-Order Harmonics} \label{gridimpacts}
\subsection{Harmonic Resonance}
\begin{figure}[h]
    \centering
\includegraphics[width=0.9\columnwidth,height=3.5in,keepaspectratio]{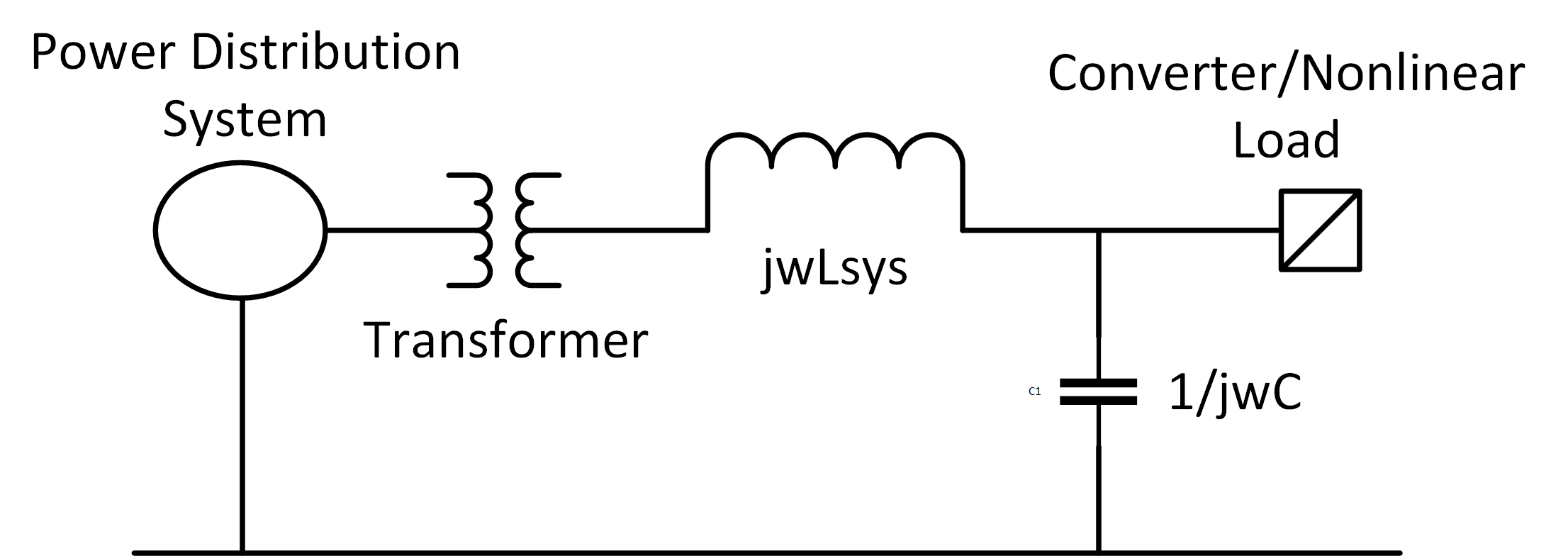}
    \caption{Simple parallel resonant circuit with capacitor bank at the nonlinear load. This case is characterized by high voltage distortion at the load and low voltage distortion upstream. }
    \label{fig:ParallelResonance}
\end{figure}

Most utilities in North America utilize three-phase or single-phase shunt-connected capacitor banks for reducing line losses and improving power factor \cite{Baghzouz1990ShuntVoltages}. However, adding capacitance to an inductive circuit creates operational conditions that can cause resonance and drastically amplify the current and voltage distortion. Parallel resonance, which is more common in distribution systems with shunt-connected capacitors, is marked by high impedance near or at resonant frequency. In parallel resonance, the current entering the parallel combination of equivalent inductance and capacitance is in phase with the supply voltage, and at the resonant frequency, there is a large circulating current between the inductive and the capacitive part of the circuit \cite{Zheng2010HarmonicFarm}. The product of large harmonic impedance and injected harmonic currents produce high harmonic voltages. Fig. \ref{fig:ParallelResonance}  shows a simple parallel resonant circuit with the shunt capacitor at the location of the nonlinear load. 
From the point of view of the nonlinear load, the power system presents a parallel combination of system inductance and capacitance, and as such the impedance seen by the nonlinear load is 
\begin{equation} \label{eq:PR}
    Z(w) = \frac{jwL_{sys}}{1-w^2L_{sys}C} 
\end{equation}
In (\ref{eq:PR}), $w=2 \pi f$ is the angular frequency measured in radians per second, $L_{sys}$ is the system inductance per phase and $C$ is the capacitance per phase. At parallel resonance, the denominator in (\ref{eq:PR}) approaches zero. However, in actual circuits, the magnitude of the impedance as seen from the load terminals is limited by the resistance of the circuit. \par
\begin{figure}[h]
    \centering
\includegraphics[width=0.9\columnwidth,height=3.5in,keepaspectratio]{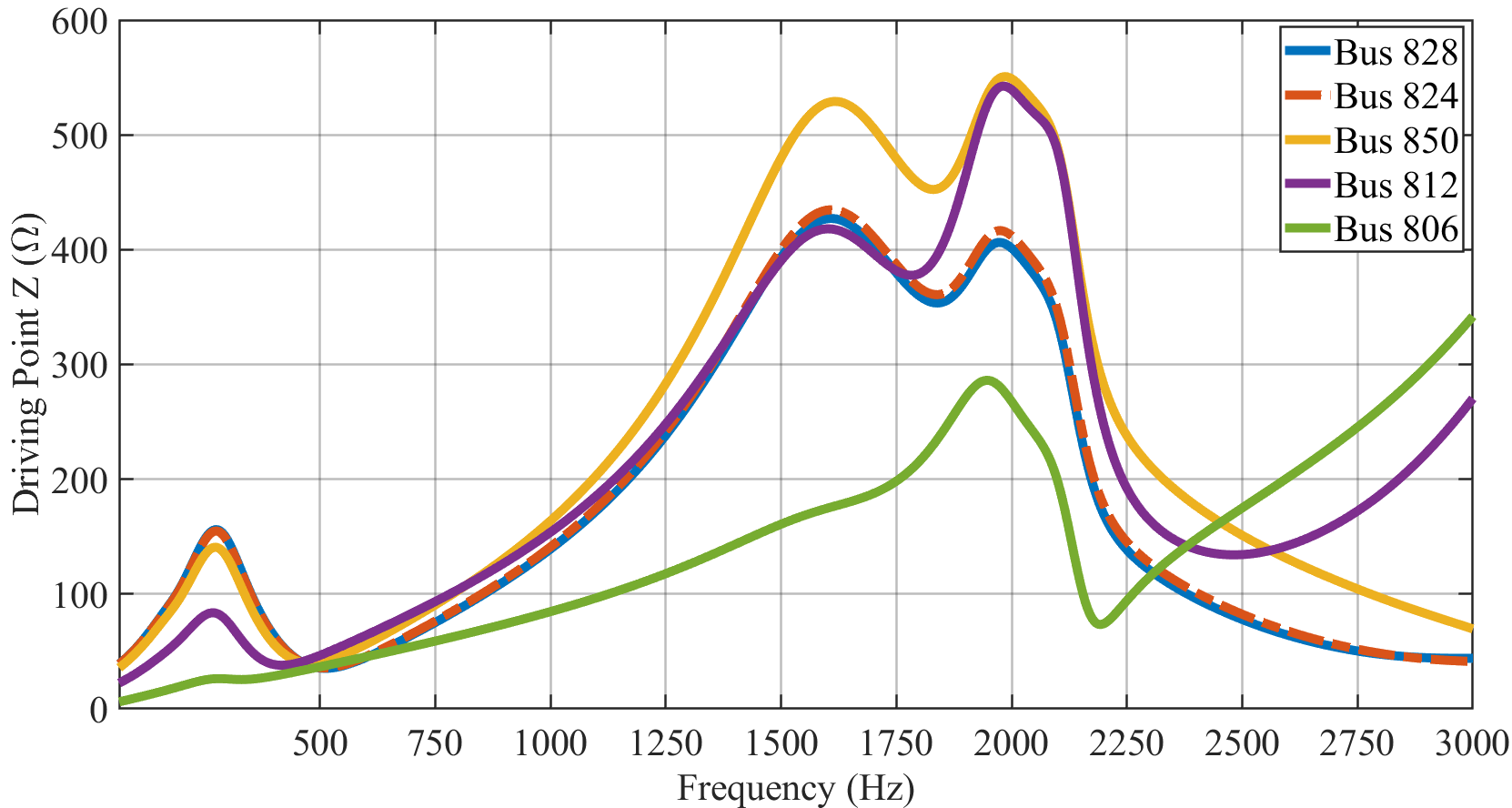}
    \caption{Harmonic resonance curves of five buses in the IEEE-34 Bus test system over a frequency range of 60 Hz-3 kHz.}
    \label{fig:Resonance}
\end{figure}

Since typical distribution feeders have more than one capacitor bank and multiple parallel and series paths, the buses in the distribution feeders are often characterized by a resonance curve i.e. a range of frequencies where local resonance may be possible. If any of the frequencies close to the peak of the resonance curve coincide with the harmonic spectrum of the nonlinear loads, it will result in high voltages and currents at those harmonics. The bus resonance curve can be obtained by running a frequency scan on the bus under consideration. This is accomplished by injecting a 1-ampere current at different frequencies and observing the corresponding voltage values. The voltages thus obtained represent the driving point impedance at the bus. An example of the variation of the driving point impedance with the frequency is shown in Fig. \ref{fig:Resonance}. The impedance curves, shown in Fig. \ref{fig:Resonance} are obtained by running a frequency scan in OpenDSS \cite{Dugan2011AnResearch} at five buses, close to the substation, in the IEEE 34 bus test system.  More details on the test system and a description of the modifications are given in Section \ref{Results}. \par
In Fig. \ref{fig:Resonance} it is worth noting that the driving point impedance at nearly all the buses reaches a peak value in the frequency range of $1.5$ kHz-$2$ kHz. At frequencies above $2.25$ kHz, the driving point impedance at buses $828$, $824$, and $850$ decreases, while at buses $812$ and $806$, it increases, indicating a higher peak above $3$ kHz. This is an important observation since modern power electronic loads are known to cause significant current and voltage waveform distortion at frequencies $\geq 1.5$ kHz \cite{Espin-Delgado2021DiagnosisEquipment}. Such high-order harmonics are common in low-voltage (LV) or medium-voltage (MV) distribution grids. If the distribution system's capacitance causes resonance at the critical high-order frequencies, it can significantly distort voltage waveform and overheat the transformer supplying the load.  
\subsection{Harmonic Power Flow}
The impact assessment of high-order harmonics on distribution grid operation is predicated on the computation of steady-state harmonic flows. In this study, we use EPRI's OpenDSS \cite{Dugan2011AnResearch} platform to solve for harmonic bus voltages. The harmonic injection currents are determined from the harmonic spectrum of the load. The harmonic spectra for different combinations of nonlinear loads (Table \ref{tab:load_models}) are obtained by running the simulation setup as shown in Fig. \ref{fig:house_combo} in PSCAD/EMTdc. The magnitude and the angle of $k^{th}$ harmonic order current are given by
\begin{equation} \label{eq:eqg1}
\begin{split}
I_{k} &= I_{1} \times K \\
\phi_k &= \phi_k + k(\phi_1-\theta)\pm 180\degree
\end{split}
\end{equation}
In (\ref{eq:eqg1}), $I_{k}$ is the magnitude of the $k^{th}$  harmonic current, $I_1$ is the magnitude of the fundamental, $K$ is the corresponding multiplier as defined in the harmonic spectrum of the nonlinear load, $\phi_k$ is the phase angle of the $k^{th}$ harmonic injection, $\phi_1$ is the phase angle of the fundamental and $\theta$ is the phase angle of the slack bus. The harmonic voltages are determined by iteratively solving 
\begin{equation} \label{eqg2}
   \bold I_k = \bold Y_k \bold V_k ; k\neq 1
\end{equation} 
In (\ref{eqg2}), $\bold Y_k$ is the bus admittance matrix with appropriate positive, negative, and zero sequence networks, $\bold I_k$ is the current injection vector at harmonic $k$, and $\bold V_k$ is the bus voltage vector at the harmonic $k$. We can rewrite (\ref{eqg2}) in terms of the slack bus and non-slack buses by partitioning the matrices as follows
\begin{equation} \label{eq:eqg3}
    \begin{bmatrix}
        I_k^{s}\\
        \bold I_k^{ns}
    \end{bmatrix}
     = \begin{bmatrix}
        \bold Y_k^{s,s} & \bold Y_k^{s,ns}\\
        \bold Y_k^{ns,s} & \bold Y_k^{ns,ns}
    \end{bmatrix}
    \begin{bmatrix}
         V_k^{s}\\
        \bold V_k^{ns}
    \end{bmatrix}
\end{equation} 
In (\ref{eq:eqg3}), $I_k^{s}$ is the current injection at the slack bus at harmonic $k$ and $V_k^{s}$ is the voltage at the slack bus at harmonic $k$. $\bold I_k^{ns}$ and $\bold V_k^{ns}$ are current injection and voltage vectors at all other buses. The voltage vector for the non-slack buses can be written in terms of the current injection vector and the partitioned system admittance matrices. The fixed-point equation for the voltages $\bold V_k^{ns}$ at iteration $m+1$ has the form
\begin{equation}
    \bold V_k^{ns, m+1} = [\bold Y_k^{ns,ns}]^{-1} (\bold I_k^{ns,m}-\bold Y_k^{ns,s} V_k^{s,m})
\end{equation}
The nodal injection current $I_k^{i,m}$ at bus $i \in \mathcal{N}$, where $\mathcal{N}$ is the set of buses, at iteration $m$ and harmonic order $k$ can be obtained from the specified real $P_i^{sp}$ and the reactive $Q_i^{sp}$ powers at bus $i$.
\begin{equation}
    I_k^{i,m} = \frac{P_i^{sp}-jQ_i^{sp}}{(V_k^{i})^{*}}
\end{equation}
\subsection{Load and DER Modeling}
\begin{figure}
    \centering
    \includegraphics[width=0.9\columnwidth,height=2.5in,keepaspectratio]{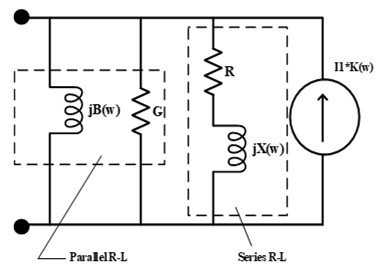}
    \caption{Nonlinear Harmonic Load Model used in the computation of harmonic flows. A $50/50$ percentage mix of the series and the parallel $R-L$ part of the load is used.}
    \label{fig:NLoad}
\end{figure}

To compute the harmonic flows, the loads on the feeder are converted into Norton Equivalents, consisting of a current source and an admittance. The nonlinear harmonic load model used to evaluate the grid impact of high-order harmonics is shown in Fig. \ref{fig:NLoad}. The model is a steady-state approximation of a VFD-based load driving an AC induction motor for HVAC application.  The shunt admittance comprises a series $R-L$ and a parallel $R-L$ part. The parameters $G$, $B$, $R$, and $L$ are obtained from the user-specified real and reactive powers of the load at $100\%$ rated voltage. 
The shunt admittance of the load plays an important role in damping out the harmonic components of the distorted load current at frequencies near or equal to the resonant frequency. This is because in the case of parallel resonance, the system impedance seen by the harmonic source increases dramatically and a significant portion of the harmonic load current is siphoned off in the shunt admittance of the load model. At frequencies away from the resonant frequency, the load impedance is much greater than the system equivalent impedance, and little current is channeled into the shunt admittance. \par 
Usually, the percentage mix of the series and the parallel R-L part of the load is unknown but in most scenarios, a $50/50$ distribution tends to produce more realistic-looking results. A $100\%$ parallel R-L load provides the highest damping and hence lowest distortion while a $100\%$ series $R-L$ load yields high values of harmonic distortion. The series $R-L$ part of the load admittance can also be used to model a rotating machine load in which case the resistance and inductance parameters are determined by blocked rotor impedance or sub-transient impedances. While it is true that the load admittance does not impact the resonant frequency in any significant manner, it does, however, provide damping to the harmonics, the amount of which depends on the type of admittance model used. The DER harmonic model representation of a PV system and an EV consists of a harmonic voltage source behind a series impedance. Similar to the load model shown in \ref{fig:NLoad}, the voltage source behind the series $R-L$ combination is initialized at the fundamental frequency using the current at the device terminals. However, in the case of the PV system and an EV, the applied spectrum represents the percentage values of voltage harmonics at different harmonic orders relative to the fundamental $60$ Hz frequency component. \par
Since a harmonic solution at each time step is preceded by a power flow solution to establish the fundamental voltages and currents, the DER models are subject to the power flow constraints. For a PV system, at each time step $t$, the real power $P^{PV}_{i,t}$ and reactive power $Q^{PV}_{i,t}$ output at bus $i\in \mathcal{N}$, under a non-unity power factor operation, is constrained to the feasible region such that $(P^{PV}_{i,t}, Q^{PV}_{i,t}) \in \mathcal{J}_{i,t} \forall i\in \mathcal{N}, t\in T$. The PV inverter feasible region is defined as
 \begin{align}
        (P^{PV}_{i,t}, Q^{PV}_{i,t}) \in \mathcal{J}_{i,t} \\
     (P^{PV}_{i,t})^2+(Q^{PV}_{i,t})^2\leq (S^{PV}_{i,t})^2\\ 
     \frac{P^{PV}_{i,t}}{\sqrt{(P^{PV}_{i,t})^2+(Q^{PV}_{i,t})^2}} \geq pf_{i,t} \label{eq:pf}
 \end{align}
 $S^{PV}_{i,t}$ is the apparent power rating of the PV inverter, and $pf_{i,t}$ in (\ref{eq:pf}) is the power factor of the $i^{th}$ PV inverter at time step $t$.  In this work, the inequality constraint of (\ref{eq:pf}) is replaced by an equality constraint considering a constant unity power factor operation of the PV system. \par  
We consider the grid-to-vehicle (G2V) operation of an EV. As such the EV power flow model calculates the power flowing into the battery module based on the power available at the point of interconnection with the rest of the distribution system. Consider $E^{+}(t)$ as the energy stored in the battery at the instant $t$. In charging mode, $E^{+}(t+\Delta t)$ is given by
\begin{equation}
    E^{+}(t + \Delta t)=E^{+}(t)+\left[\eta_{inv}(t) P_{in}(t)-P_{idl}\right]\eta_{ch} \Delta t  \label{eq:BESS}
\end{equation}
In (\ref{eq:BESS}), $\eta_{inv}(t)$ is the inverter efficiency at time $t$, $\eta_{ch}$ is the inverter charging efficiency, $P_{in}(t)$ is the real power injected into the battery storage from the grid at $t$, and $P_{idl}$ is the constant idling loss. The power flow constraints are imposed by the battery capacity limits and the state of charge (SoC). 
\begin{align}
     E_{max}^{-}\leq E_{BESS}^{+}(t)\leq E_{max}^{+}\\
     SoC_{min}\leq SoC_{BESS}(t)\leq SoC_{max} \label{eq:BESS3}
 \end{align}
In (\ref{eq:BESS3}), $E_{max}^{+}$ and $E_{max}^{-}$ are the maximum energy capacities in charging and discharging mode respectively. $ SoC_{min}$, $SoC_{max}$ and $SoC_{BESS}(t)$ represent the minimum, maximum, and current battery state of charge. \par 
In OpenDSS, the harmonic models of PV systems and EVs consist of a voltage harmonic source behind a series impedance. The initialization of the voltage source behind the series impedance is performed at the fundamental frequency by considering the current in the terminal. The defined spectrum is applied to the voltage source, as opposed to the nonlinear load, where the defined spectrum is applied relative to the current source. 

\section{Results} \label{Results}

\subsection{Simulation Setup}
\begin{figure}
    \centering
    \includegraphics[width=0.9\columnwidth,height=2in,keepaspectratio]{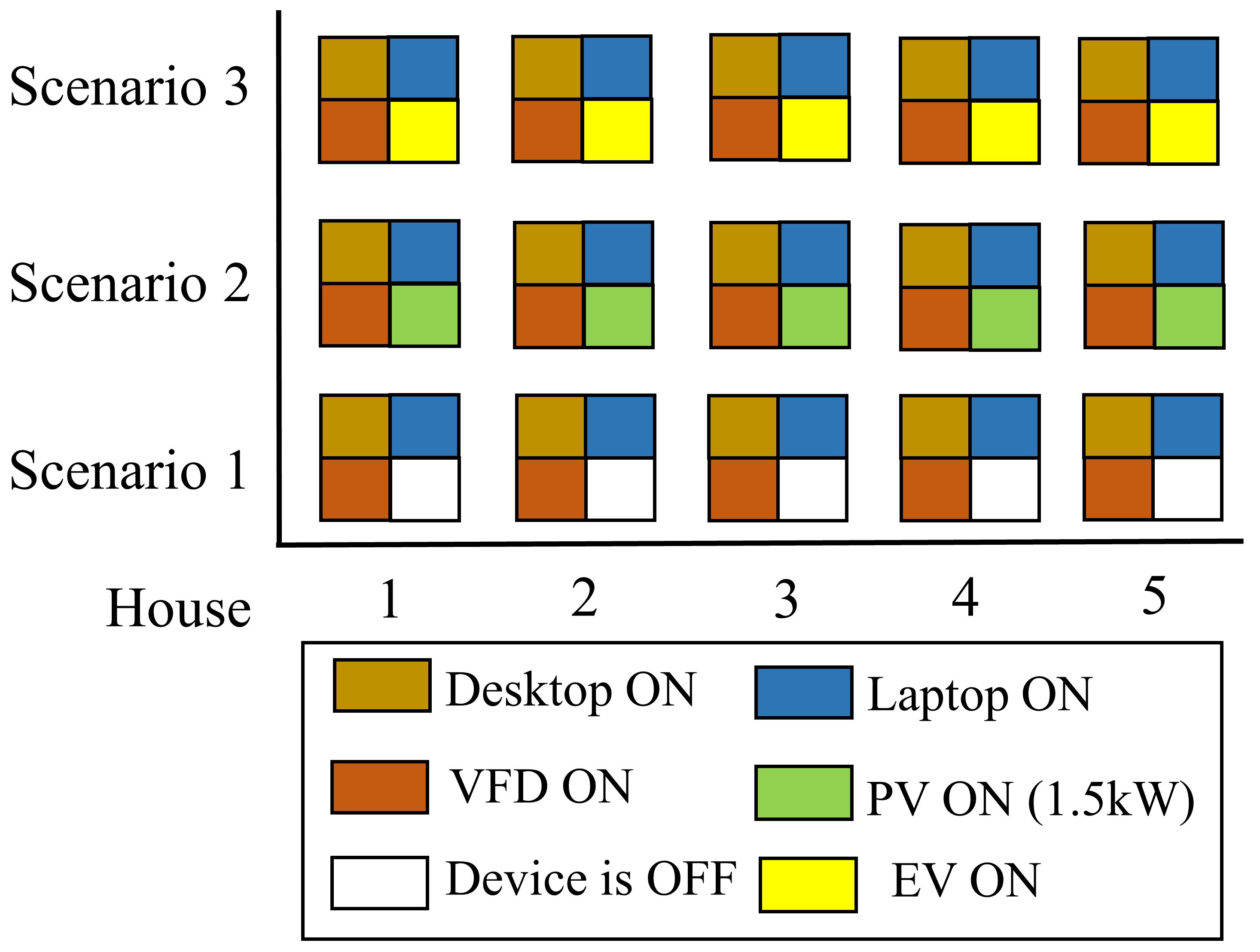}
    \caption{Load combination for 5 houses corresponding to a different transformer loading condition.}
    \label{fig:cases}
\end{figure}

To understand the effect of different loading scenarios on transformers,  the simulation setup described in Fig. \ref{fig:cases} is used. A total of 3 scenarios are constructed representing different penetration levels of electronic components. This helps us to analyze the impact of \gls{pv}, \gls{ev}, and \gls{vfd} penetration on a residential distribution transformer, as shown in Table \ref{tab:load_models}. All these scenarios are assumed to represent the peak loading condition for a given transformer. These scenarios replicate different loading periods throughout the day, where either \gls{pv}, \gls{ev}, or \gls{vfd} are the major contributors to residential loads. \par 

% \begin{figure}
%     \centering
%     \includegraphics[width=1\columnwidth]{Figures/transformer_sec.png}
%     \caption{Load combination for 5 houses corresponding to a different transformer loading condition.}
%     \label{fig:cases}
% \end{figure}

\begin{figure}[h]
    \centering
  \includegraphics[width=0.9\columnwidth,height=3.5in,keepaspectratio]{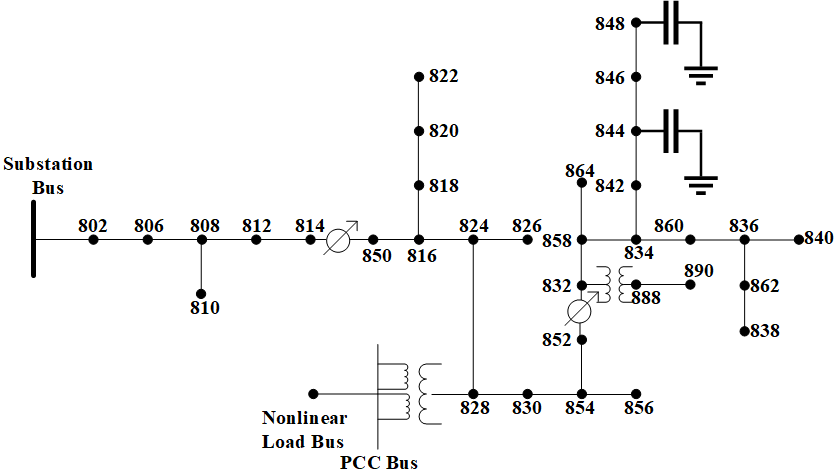}
    \caption{Modified IEEE-34 Bus Test System with a 50 kVA split-phase center tapped transformer used for interconnecting the aggregate nonlinear load with the distribution system.}
    \label{fig:test_system}
\end{figure}

% \begin{figure}
%     \centering
%     \includegraphics[width=1\columnwidth]{Figures/thd_eddy_loss.png}
%     \caption{THD measures.}
%     \label{fig:cases}
% \end{figure}
To quantify the grid impact of high-order harmonics, we consider the load that is an aggregate of the five houses shown in Fig. \ref{fig:cases}. Specifically, we consider three operational scenarios with variations in the composition of the nonlinear load. In scenario 1, the aggregated nonlinear load is assumed to consist of a \gls{vfd}, a desktop, and a laptop. In scenario 2, a \gls{pv} system is added to the load composition used in scenario 1. In scenario 3, an \gls{ev} is added to the nonlinear load in scenario 1. For all the scenarios, the aggregated harmonic spectrum is obtained from the recorded current and voltage data at the transformer secondary sampled at a frequency of $20$ kHz. The grid impact assessment is performed using an equivalent steady-state model of the simulation setup shown in Fig. \ref{fig:house_combo} and interfacing it with a model of the IEEE-34 bus test system. The modified IEEE-34 bus test feeder is shown in Fig. \ref{fig:test_system}. 

A $50$ kVA single-phase, three-winding center-tapped transformer with a primary leakage impedance of $2.04\%$ in delta/wye-wye is used to interconnect the aggregate nonlinear load with the IEEE-34 distribution system. The secondary line-ground (L-G) voltages across the split phase are $0.12$ kV and $0.207$ kV when measured across line-line (L-L). To avoid greatly exaggerating the voltage distortion at the point of common coupling (PCC), the apparent resistance of the transformer is increased proportionally with the reactance as the frequency increases. This adjustment allows for a constant $X/R$ ratio of the transformer over a wide range of frequencies. \par
\begin{figure}[h]
    \centering
  \includegraphics[width=0.9\columnwidth,height=3.5in,keepaspectratio]{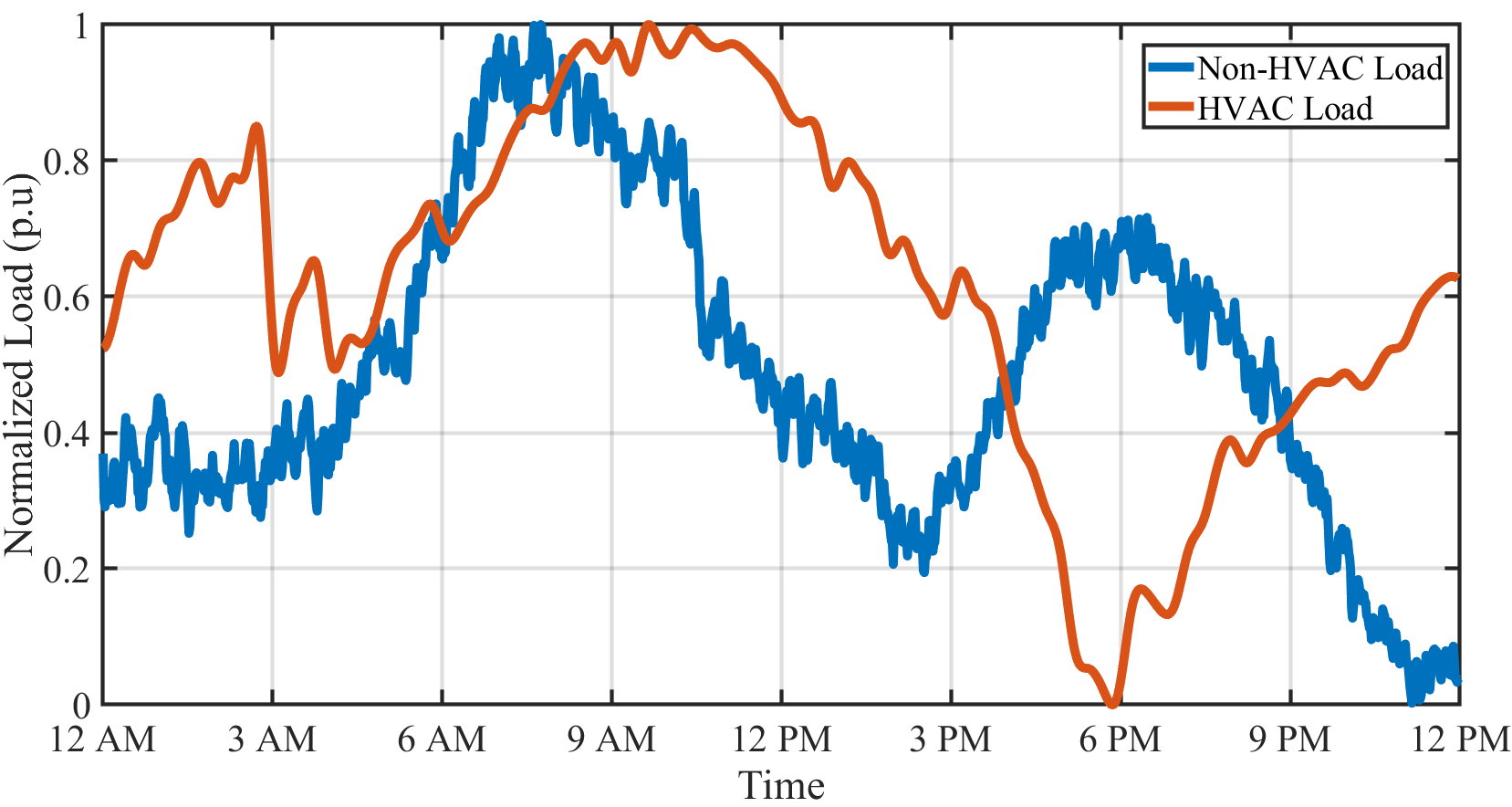}
    \caption{Daily (24-hour) variation in the linear and the nonlinear aggregate residential load.}
    \label{fig:load}
\end{figure}

 
The aggregated load (representing all five houses) is split into a linear and a nonlinear part with appropriate load profiles as given in \cite{osti_1414819}. The daily (24-hour) linear and nonlinear load profiles used to control aggregate load power injection are shown in Fig. \ref{fig:load}. \par
\begin{figure}[h]
    \centering
  \includegraphics[width=0.9\columnwidth,height=3.5in,keepaspectratio]{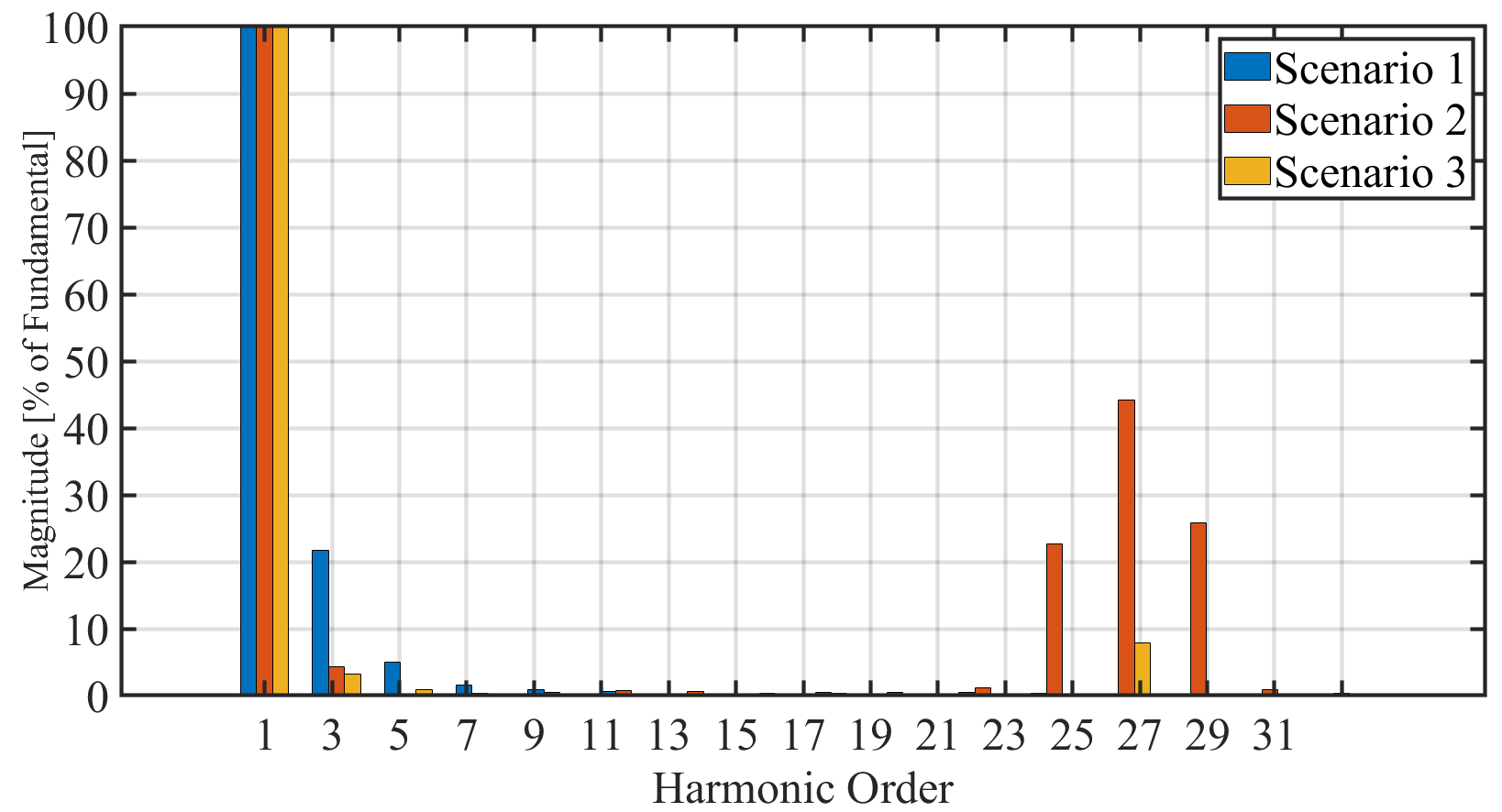}
    \caption{Current Harmonic Spectrum of different scenarios obtained from PSCAD/EMTdc simulation.}
    \label{fig:spectra}
\end{figure}

The impact of each scenario as outlined in Fig. \ref{fig:cases} on the grid operation is evaluated using a unique current and voltage harmonic spectrum derived from the PSCAD/EMTdc simulation. Fig. \ref{fig:spectra} shows the aggregate generated harmonic spectrum of three scenarios. It is important to note that only scenario 2 exhibits predominant harmonics at and close to the $27^{th}$ harmonic order. This corresponds to a frequency of $\approx 1.6$ kHz and aligns perfectly well with the resonant frequency at certain locations on the IEEE-34 test feeder as demonstrated in Fig. \ref{fig:Resonance}. Due to this alignment, the distortion is expected to be very high in scenario 2 compared to scenarios 1 and 3. 
\subsection{THD and Transformer Eddy Loss}
\begin{figure}[h]
    \centering
  \includegraphics[width=0.9\columnwidth,height=3.5in,keepaspectratio]{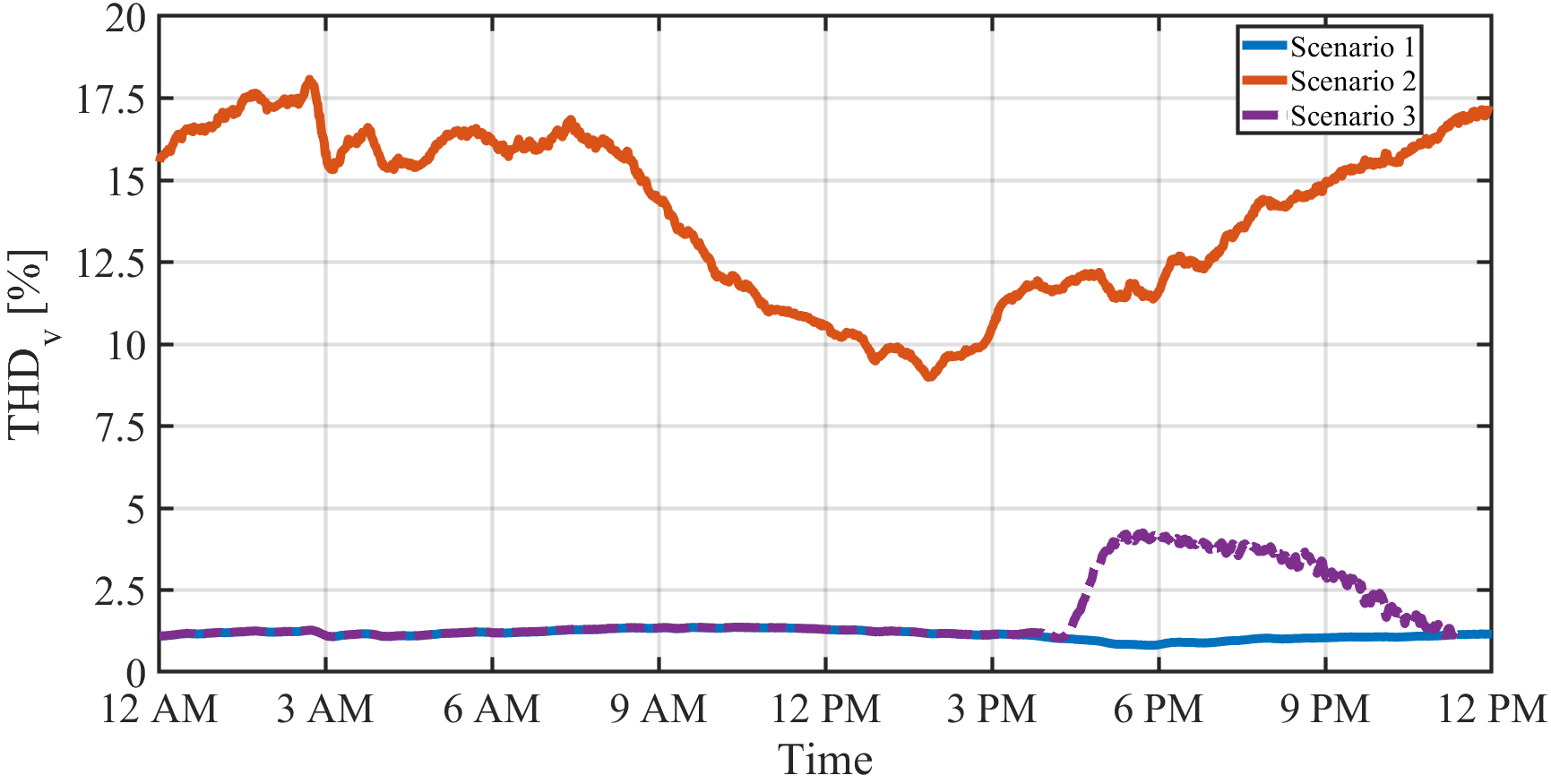}
    \caption{Daily (24-hour) voltage THD profiles at the PCC bus for three scenarios. The distortion is very high in scenario 2 because of the harmonics injected at and close to the $27^{th}$ harmonic which is very close to the resonant frequency at the PCC bus.}
    \label{fig:thd_time}
\end{figure}

\begin{figure}[h]
    \centering
  \includegraphics[width=0.9\columnwidth,height=3.5in,keepaspectratio]{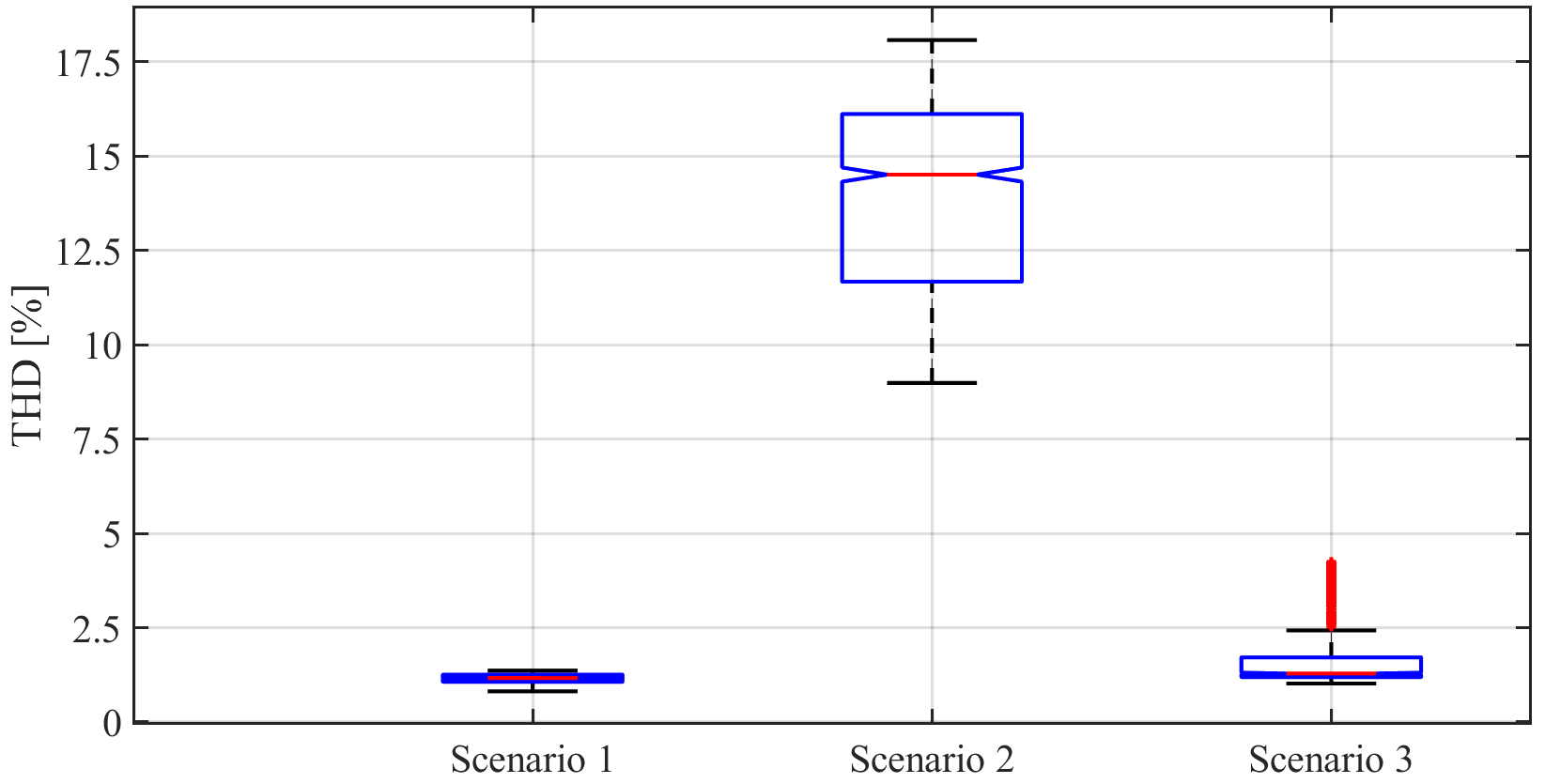}
    \caption{Box plot showing the range of voltage THD variation at the PCC bus for three scenarios. The THD is highest (peak and average) for Scenario 2.}
    \label{fig:THD}
\end{figure}

With the aggregated nonlinear load at the PCC bus, the time-series harmonic solutions for each scenario are obtained using the sub-hourly (1-minute resolution) load profiles and the associated harmonic spectrum for each scenario. It is important to note that the resonant frequency at the PCC bus is $1.5$ kHz, representing the $25^{th}$ harmonic order. The time series voltage THD profile at the PCC bus for three scenarios is shown in Fig. \ref{fig:thd_time}. Fig. \ref{fig:THD} shows the range of THD in bus voltage and the average values for the three scenarios observed at the PCC bus.

The significant voltage distortion in scenario 2 directly results from the interaction between the resonance curve at the \gls{pcc} bus and the harmonic spectrum of the load. Unlike scenario 1 and scenario 3, the spectrum of scenario 2 consists of predominant $25^{th}$, $27^{th}$, and $29^{th}$ harmonics and these are either very close or equal to the resonant frequency at the \gls{pcc} bus. This alignment of injected harmonics and feeder resonance amplifies the voltage distortion and may cause significant transformer overheating. Furthermore, in scenario 3, the voltage THD increases sometime past 3 PM, reaching a peak value close to 6 PM, and starts to decrease after that until it is equal to the THD value in scenario 1. This is due to the EV charging which begins after 3 PM. The charging stops when the battery SoC reaches $95$ of the battery capacity. \par
\begin{figure}[h]
    \centering
  \includegraphics[width=0.9\columnwidth,height=3.5in,keepaspectratio]{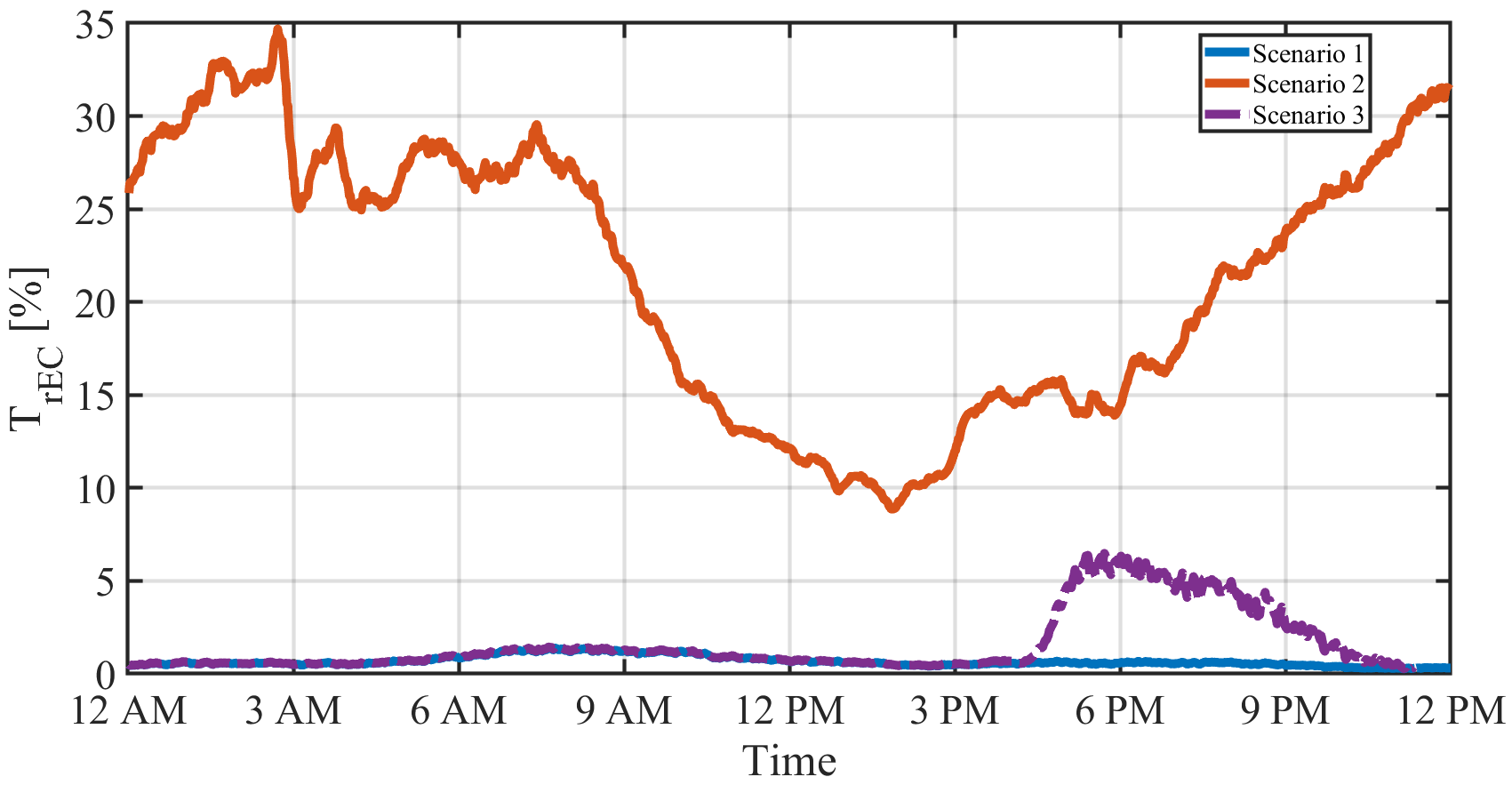}
    \caption{Daily (24-hour) variation in the harmonic-driven eddy-current loss component in the split-phase center-tapped transformer. The eddy current loss is highest, on average as well as the peak value, for scenario 2.}
    \label{fig:eddy_time}
\end{figure}

\begin{figure}[h]
    \centering
  \includegraphics[width=0.9\columnwidth,height=3.5in,keepaspectratio]{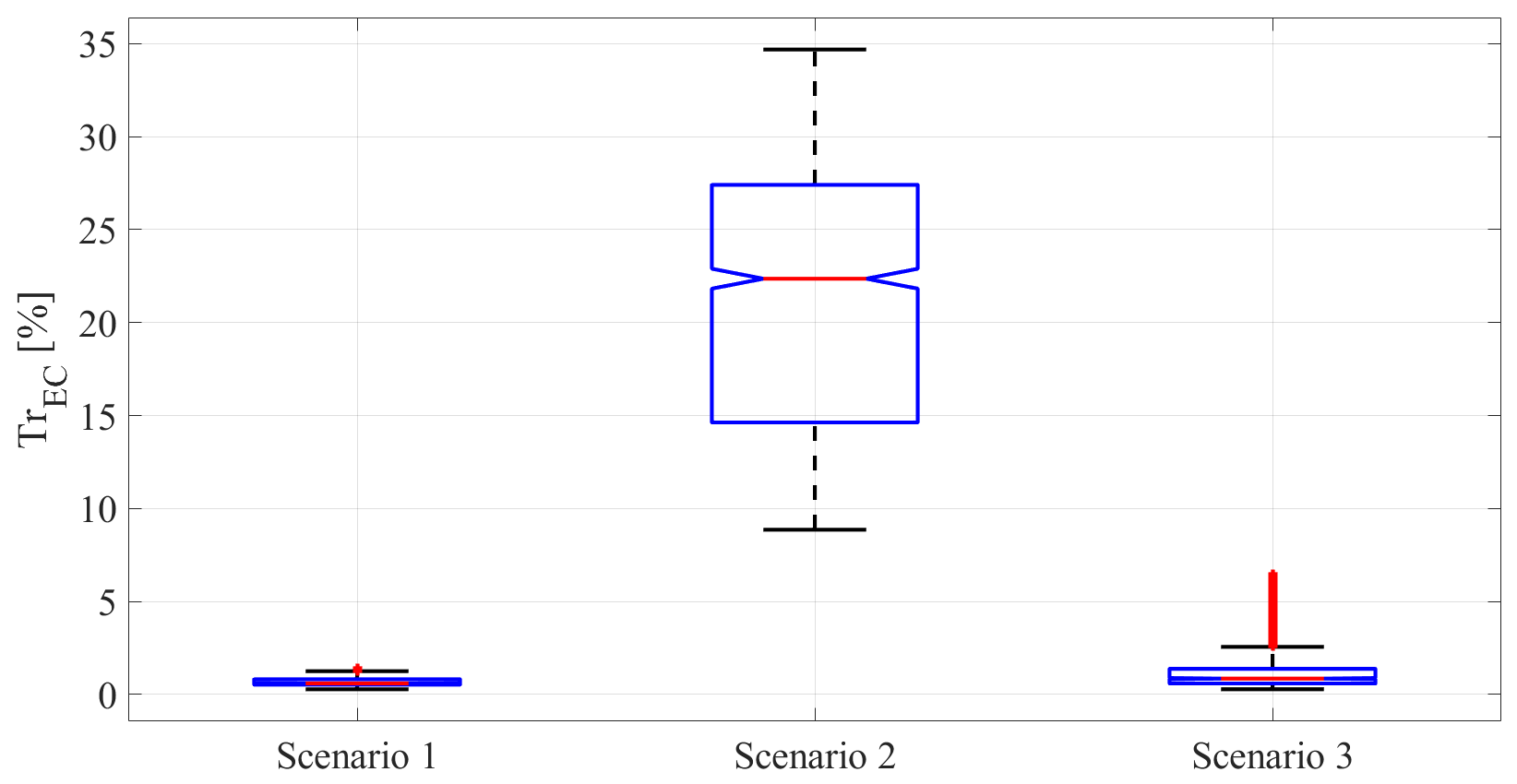}
    \caption{Box plot showing the range of harmonic-driven eddy current loss component. The eddy current loss is highest for scenario 2. }
    \label{fig:EddyLoss}
\end{figure}

 
Fig. \ref{fig:eddy_time} and Fig. \ref{fig:EddyLoss} show the variation in the harmonic-drive eddy current loss component and the range of the harmonic-driven transformer eddy current loss component, respectively, of the split-phase center-tapped transformer.

The harmonic-driven eddy current loss component is given by
\begin{equation} \label{eq:R1}
    Tr_{EC} = P_{EC-R} \sum_{h=1}^{h_{max}} I_h^{2}h^2
\end{equation}
In (\ref{eq:R1}), $P_{EC-R}$ is the winding eddy current loss factor and is assumed to be $0.05$. More details on the calculation of the harmonic-drive eddy current loss component are given \cite{mitra2023}. 
\subsection{\gls{thd} Propagation Upstream}
\begin{figure}[h]
    \centering
  \includegraphics[width=0.9\columnwidth,height=3.5in,keepaspectratio]{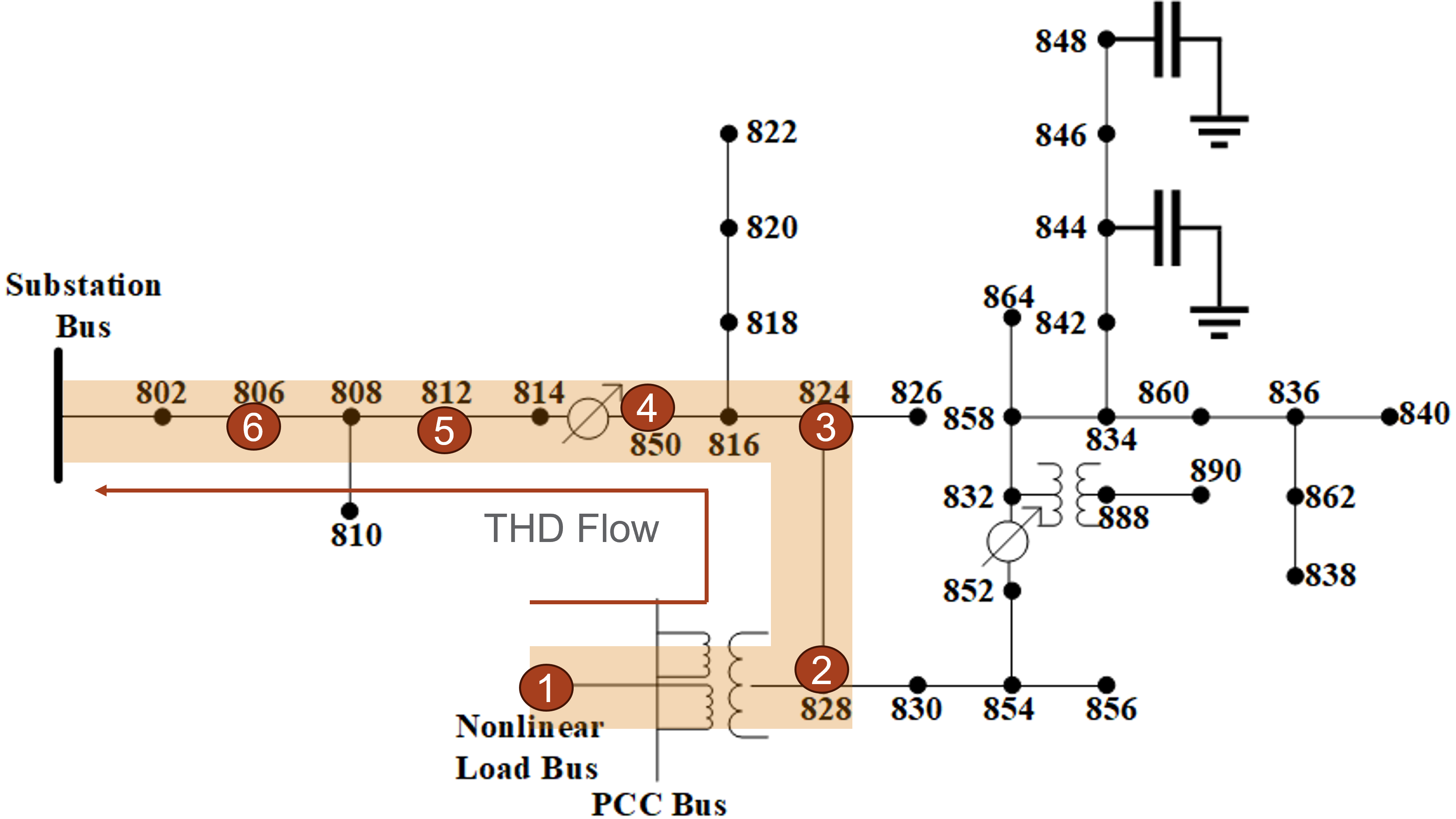}
    \caption{Sequential placement of nonlinear loads on the IEEE-34 radial network at locations with resonant frequency in the range of $1.5$ kHz-$2$ kHz as shown in Fig. \ref{fig:Resonance}.}
    \label{fig:Feeder_new}
\end{figure}

Under normal operating conditions, the electric distribution systems are built with sufficient kVA capacity to supply the kVA load while maintaining the voltage distortion levels within the acceptable range. However, the shunt capacitor banks and the system inductance can cause feeder resonance and magnify the distortion beyond acceptable limits. Moreover, if there are multiple dispersed sources of harmonic injection with critical harmonic frequencies close to or equal to the resonant frequency at a specified location, the overall voltage distortion will increase drastically and may propagate upstream to the substation bus in a radial network. To observe the \gls{thd} propagation in a radian network, we place nonlinear loads sequentially at locations where the resonant frequency is equal to or close to one of the predominant critical harmonic frequencies in the loads. The nonlinear loads use a harmonic spectrum for scenario 2 since that results in the highest \gls{thd} distortion given the presence of predominant $25^{th}$, $27^{th}$, and $29^{th}$ harmonic orders. Fig. \ref{fig:Feeder_new} shows the sequential placement of nonlinear loads along the length of the feeder. 

\begin{figure}[h]
    \centering
  \includegraphics[width=0.9\columnwidth,height=3.5in,keepaspectratio]{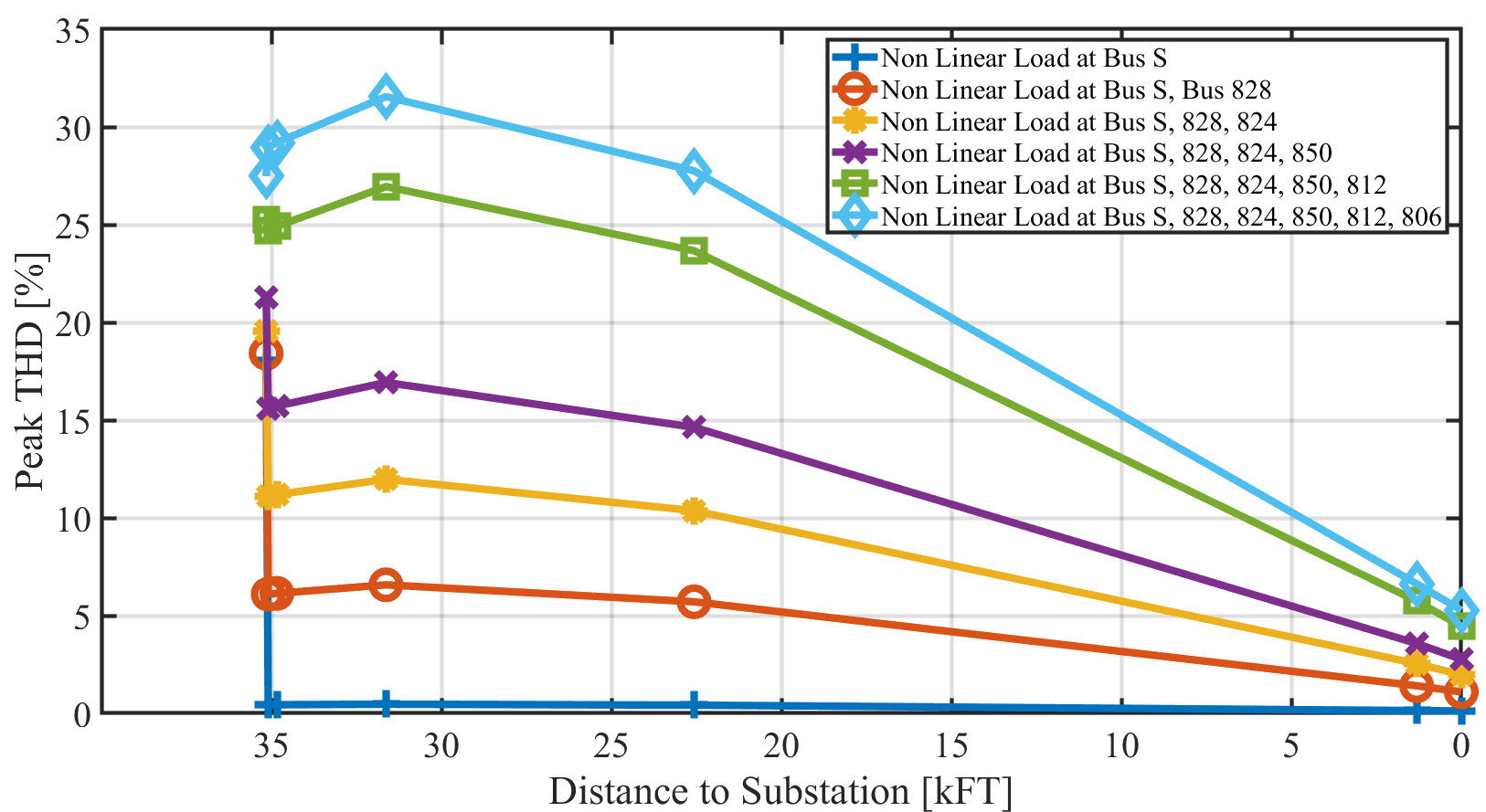}
    \caption{Peak THD propagation along the feeder with sequential placement of nonlinear loads. Note that Bus S in the legend refers to the nonlinear load bus in Fig. \ref{fig:Feeder_new}.}
    \label{fig:THD_prop}
\end{figure}

The THD propagation is measured by placing monitors at the locations highlighted in Fig. \ref{fig:Feeder_new} and running time-series harmonic simulations with appropriate load profiles and harmonic spectra. The voltage harmonics at each location are recorded at every time step and the \gls{thd} is computed. Fig. \ref{fig:THD_prop} shows the peak THD propagation with sequential placement of nonlinear loads. 

\begin{figure}[h]
    \centering
  \includegraphics[width=0.9\columnwidth,height=3.5in,keepaspectratio]{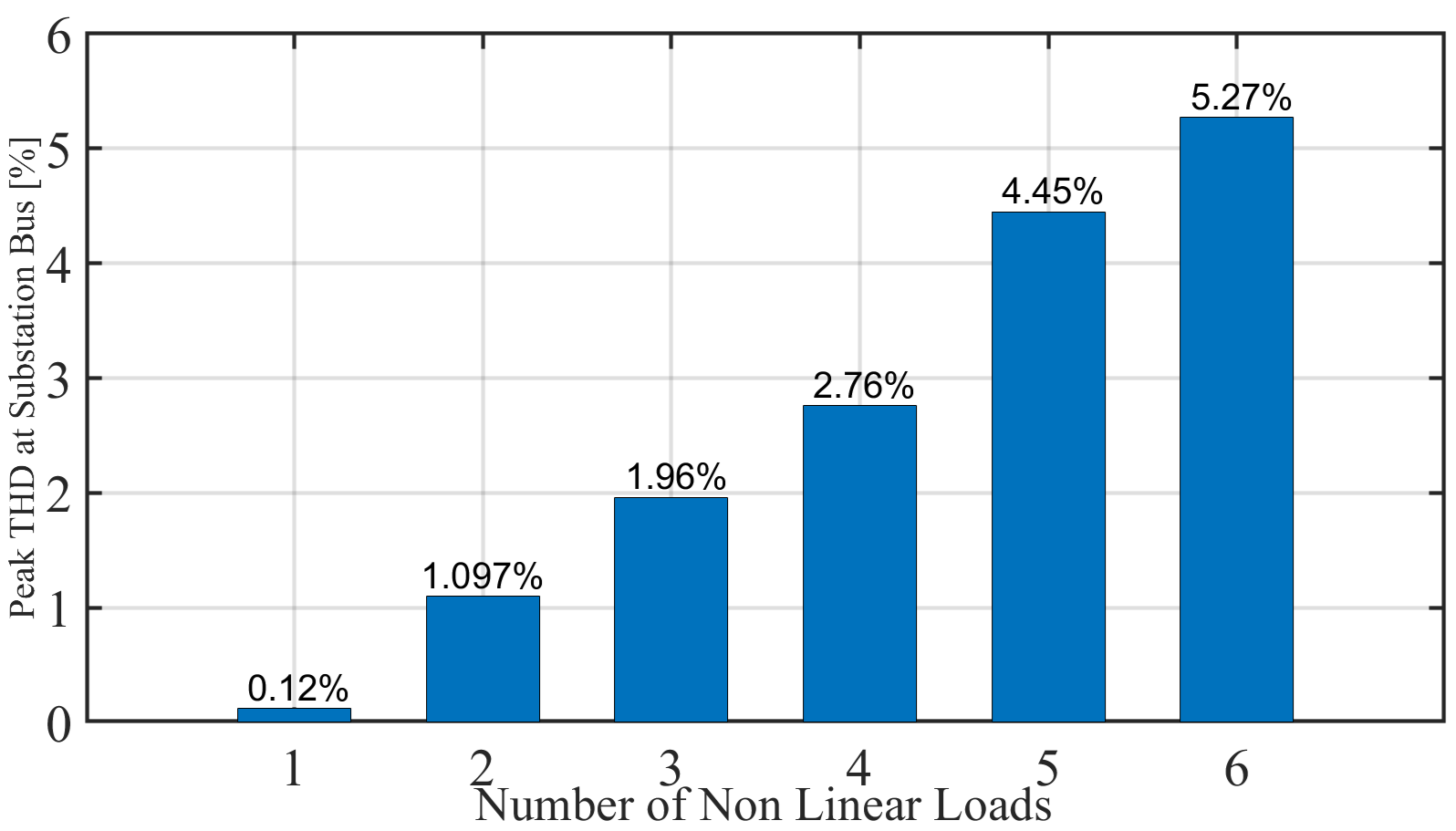}
    \caption{Peak \gls{thd} measured at the substation bus with the increasing number of nonlinear loads.}
    \label{fig:THD_source}
\end{figure}

The peak value of the \gls{thd} at each bus is computed from the $24$-hour time-series \gls{thd} values and it is clear from Fig. \ref{fig:THD_prop} the peak \gls{thd} at each bus increases with the sequential placement of nonlinear loads. As more nonlinear loads are connected at locations that allow for interference between the resonant frequency and the injected harmonics, the peak \gls{thd} measured at the substation bus increases. The effect of damping offered by the resistance of the cables is also evident in Fig. \ref{fig:THD_prop}. The \gls{thd} propagation curves as shown in Fig. \ref{fig:THD_prop} are characterized by a decrease in the peak \gls{thd} level as the distance to the substation bus decreases;  a consequence of the damping of the resonance by the cable resistance. The change in the peak \gls{thd} measured at the substation bus with the increase in the number of nonlinear loads is shown in Fig. \ref{fig:THD_source}.

The substation \gls{thd} exceeds the $5\%$ threshold mark with six nonlinear loads each comprising of a \gls{vfd}, a desktop computer, a laptop computer, and a \gls{pv} system. This underscores the damaging effect of the constructive interference between the system resonance and the injected higher-order harmonics.
\section{Conclusion} \label{conc}
In this work, we present an empirical study to demonstrate the impact of high-order harmonics on power quality in electrical distribution systems. More specifically, we present empirical evidence to support the claim that the feeder resonance, which results from adding capacitance to an inductive circuit, can significantly amplify the current and the voltage distortion whenever the injected harmonics are equal to or close to the bus resonant frequency. This fact is demonstrated by calculating the \gls{thd} values, using time-series harmonic simulations, at the point of interconnection of the nonlinear load and observing the change in the peak \gls{thd} values at the substation bus when multiple dispersed sources of harmonic injection are considered. Moreover, the effect of high-order harmonics on the transformer heating is illustrated by calculating the time-series harmonic-driven eddy current loss component. The impact of high-order harmonics is quantified by developing three operational scenarios of different combinations of nonlinear loads. It is shown that only scenario 2 injects high-order harmonics and as a result is characterized by the highest average and peak \gls{thd} and higher average and peak eddy current loss.
    %\nocite{*}
    \bibliographystyle{ieeetr}
    \bibliography{references.bib}
    %\bibliography{references_2.bib}
\end{document}